\documentclass[prd,%
reprint, 
amsmath,amssymb,
aps,
]{revtex4-2}

\usepackage{graphicx}
\usepackage{dcolumn}
\usepackage{bm}
\usepackage{hyperref}
\usepackage{color}

\begin{document}

\title{Timelike Entanglement Entropy of Hawking Radiation }

\author{Yahya Ladghami$^{1,2}$}
\email{yahya.ladghami@ump.ac.ma}

\author{Francisco S.N. Lobo$^{3,4}$}
\email{fslobo@ciencias.ulisboa.pt}
	
\author{Taoufik Ouali$^{1,2}$}
\email{t.ouali@ump.ac.ma}

\affiliation{$^1$Laboratory of Physics of Matter and Radiation, Mohammed I University, BP 717, Oujda, Morocco}
\affiliation{$^2$Astrophysical and Cosmological Center, BP 717, Oujda, Morocco}
\affiliation{$^3$	Instituto de Astrof\'{\i}sica e Ci\^encias do Espa\c{c}o, Faculdade de Ci\^encias da Universidade de Lisboa,
Edif\'{\i}cio C8, Campo Grande, P-1749-016 Lisboa, Portugal}
\affiliation{$^4$ Departamento de F\'{\i}sica, Faculdade de Ci\^encias da Universidade de Lisboa,\\
Edif\'{\i}cio C8, Campo Grande, P-1749-016 Lisboa, Portugal
		 }

\date{\today}

\begin{abstract}
	We introduce the concept of timelike entanglement entropy of Hawking radiation as a novel probe of the black hole information paradox. By analytically continuing black hole spacetimes to Euclidean signature, we define timelike correlations that reveal a sequence of timelike Page times at which the entanglement entropy equals the Bekenstein-Hawking entropy. Applying this framework to Schwarzschild, Reissner-Nordstr\"om, static higher-dimensional and braneworld solutions, four-dimensional Kerr, and higher-dimensional rotating Myers--Perry black holes, we demonstrate that timelike entanglement exhibits periodic or quasi-periodic behavior, with the recurrence times  sensitive to surface gravity, charge, rotation, and spacetime dimensionality. Extremal and near-extremal black holes display effectively frozen thermal oscillations with persistent rotational modulation, reflecting their near-horizon geometries. Unlike conventional approaches based on islands or firewalls, our framework encodes information entirely in the Hawking radiation, preserving unitarity while avoiding violations of horizon smoothness. These results establish timelike entanglement as a robust and physically transparent mechanism for information recovery and provide a versatile tool for exploring quantum gravitational dynamics across a wide range of black hole spacetimes.     
\end{abstract}

\maketitle


\section{Introduction}

Black holes occupy a central position in modern theoretical physics, as they constitute a unique arena in which gravity, quantum mechanics, and thermodynamics intersect. Beyond their astrophysical relevance, black holes serve as a powerful theoretical laboratory for the development and testing of candidate theories of quantum gravity, including loop quantum gravity \cite{RovelliLQG}, string theory \cite{PolchinskiBook}, and holographic duality \cite{Maldacena1997}. As such, they are widely regarded as a cornerstone for understanding the fundamental nature of spacetime at the quantum level.

A distinctive feature of black hole physics is black hole evaporation: Hawking showed that black holes emit thermal radiation with a blackbody spectrum, now known as Hawking radiation \cite{Hawking1975}, causing an isolated black hole to lose mass and potentially evaporate completely. This process, however, leads to a fundamental tension with quantum mechanics. While a black hole may form from matter in a pure quantum state, Hawking radiation is thermal and is therefore described by a mixed state \cite{Hawking1976}. Complete evaporation would therefore imply an evolution from a pure to a mixed state, violating quantum mechanical unitarity. This apparent violation is manifested in the monotonically increasing von Neumann entropy of the emitted radiation in Hawking's semiclassical description and is known as the black hole information loss paradox.

In an effort to reconcile black hole evaporation with quantum unitarity, Page proposed a characteristic behavior for the entanglement entropy of Hawking radiation, now referred to as the Page curve \cite{Page1993a,Page1993b}. According to this proposal, the entropy of the radiation initially increases, reaches a maximum at the so-called Page time, and subsequently decreases as the black hole continues to evaporate. This late-time decrease implies that information encoded in the initial quantum state is gradually recovered in the radiation, preserving unitarity.

Several theoretical mechanisms have been proposed to explain the emergence of the Page curve within a consistent framework of quantum gravity. Two prominent approaches are the firewall proposal \cite{AMPS2013}, which proposes a breakdown of effective field theory at the event horizon, and the island formula \cite{Penington2019,Almheiri2019}, which modifies the semiclassical calculation of entropy by including additional contributions from quantum extremal surfaces. These approaches represent distinct attempts to resolve the information loss paradox and remain central to ongoing research in black hole physics.

The firewall proposal was introduced as a possible resolution of the black hole information loss paradox under the assumption that quantum mechanical unitarity must be preserved \cite{AMPS2013}. In classical general relativity, the event horizon of a black hole is locally smooth, and an observer freely falling across the horizon experiences no anomalous physical effects, in accordance with the equivalence principle. The firewall proposal challenges this expectation by arguing that such smoothness cannot be maintained if unitarity is required.

However, before the Page time, Hawking radiation is predominantly entangled with the remaining black hole, leading to an increase in the entanglement entropy of the radiation. After the Page time, however, unitarity demands that newly emitted Hawking quanta be entangled with the early radiation rather than with the black hole interior, so that information about the initial quantum state can be recovered in the outgoing radiation \cite{Page1993a}. At the same time, the equivalence principle implies that outgoing modes near the horizon must be entangled with their interior partners in order to reproduce the local vacuum state experienced by an infalling observer.

These two entanglement requirements are mutually incompatible due to the monogamy of entanglement, which forbids a quantum system from being simultaneously maximally entangled with two independent systems \cite{Coffman2000}. As a result, the entanglement between outgoing Hawking radiation and the black hole interior must be disrupted. The firewall proposal posits that this disruption manifests as a highly energetic region, or ``firewall,'' located at the event horizon, which destroys the interior entanglement and leads to violent interactions for any observer attempting to cross the horizon after the Page time \cite{AMPS2013}. In this way, the firewall scenario preserves quantum unitarity and reproduces the Page curve, but does so at the cost of violating the equivalence principle and the smoothness of spacetime at the horizon \cite{AMPS2013}.

An alternative method for addressing the black hole information problem is provided by the island formula~\cite{i,Almheiri2019}. Within this framework, the entanglement entropy of Hawking radiation exhibits two distinct regimes. Prior to the Page time, the entropy is computed exclusively from the degrees of freedom of the radiation itself, leading to a monotonically increasing entanglement entropy as the black hole evaporates. 
After the Page time, however, a new contribution becomes relevant: a spacetime region located inside the black hole, known as the \emph{island}, dynamically emerges~\cite{i}. In this late-time regime, the entanglement entropy is evaluated by including contributions from both the Hawking radiation and the island region. The location and extent of the island are determined through the gravitational path integral using replica wormhole configurations. This prescription successfully reproduces the Page curve and yields an entropy evolution that is consistent with unitary quantum evolution~\cite{iii}.

Despite these notable successes, the island formula is subject to several important limitations. At present, the framework is rigorously established only within the context of the AdS/CFT correspondence, while its applicability to asymptotically flat or de Sitter spacetimes remains unclear~\cite{ss,sss,ssss}. In addition, certain analyses indicate that island formation may not be a universal feature of all black hole solutions; for example, islands appear to be absent in Liouville black holes~\cite{Lii}. Finally, the physical interpretation of islands is still not well understood, and the formalism does not yet provide a clear microscopic mechanism by which information is recovered from the black hole.

Within the broader study of quantum entanglement, a novel notion has been introduced, known as temporal or timelike entanglement~\cite{Oo}. This concept characterizes quantum correlations between states of the same physical system evaluated at different moments in time, rather than at different spatial locations~\cite{AA}. This stands in clear contrast to classical correlations or conventional spatial quantum entanglement, which describe correlations between distinct subsystems that are separated in space.
Timelike entanglement can be systematically analyzed using several theoretical frameworks, including the Choi-Jamiołkowski isomorphism~\cite{DD,EE,J3}, which maps quantum channels to quantum states, as well as the Feynman-Vernon influence functional formalism~\cite{FF}, which captures the effects of temporal correlations in open quantum systems. More recently, this idea has been extended to the context of the AdS/CFT correspondence, giving rise to the concept of \emph{holographic timelike entanglement entropy}~\cite{H3}. This quantity may be viewed as a temporal generalization of the Ryu-Takayanagi formula.

Unlike the standard Ryu-Takayanagi prescription~\cite{r}, which computes entanglement entropy for spatial subregions defined on a single time slice, timelike entanglement entropy applies to subregions that are separated along the time direction. As a consequence of this temporal partitioning, the associated density matrix is generally non-Hermitian, and the resulting holographic timelike entanglement entropy is therefore a complex-valued quantity. This entropy can be evaluated through an appropriate Wick rotation and is naturally interpreted as a \emph{pseudoentropy}~\cite{t}, which generalizes the von Neumann entropy and provides a measure of quantum correlations between two distinct quantum states.
One of the most significant implications of this framework is the suggestion that time itself may emerge from underlying quantum entanglement, in a manner analogous to the emergence of space from spatial entanglement as captured by the Ryu-Takayanagi formula~\cite{r}.

Thus, the goal of this work is to examine whether the black hole information paradox can be addressed from a new perspective based on timelike quantum correlations. Specifically, we investigate the role of \emph{timelike entanglement entropy} of Hawking radiation as a probe of information retention and recovery during black hole evaporation.
Instead of focusing solely on spatial entanglement between the black hole interior and outgoing radiation, we analyze quantum correlations among radiation states emitted at different times. This approach allows us to assess whether information about the initial black hole state can be encoded in temporal correlations and whether such correlations admit a consistent unitary description of the evaporation process.
Our framework is complementary to existing proposals, such as the island formula, as it replaces spatial partitions of spacetime with temporal partitions of quantum states. Through this shift in perspective, we aim to determine whether timelike entanglement entropy can capture essential features of unitary black hole evaporation and provide insight into the microscopic origin of information preservation.

The paper is organized as follows. In Sec.~\ref{SecII}, we introduce holographic timelike entanglement entropy and the framework for analyzing temporal correlations in Hawking radiation. Sec.~\ref{SecIII} applies this framework to various black holes, including Reissner-Nordstr\"om, static higher-dimensional and braneworld solutions, four-dimensional Kerr, and higher-dimensional rotating Myers-Perry black holes, highlighting the effects of charge, rotation, dimensionality, and extra-dimensional corrections on timelike Page times and information recovery. Finally, Sec.~\ref{Sec:Conclusion} summarizes our results, discusses implications for the black hole information paradox, and outlines directions for future research.

\section{Holographic Timelike Entanglement Entropy}\label{SecII}

The AdS/CFT correspondence leads to the idea that the spatial direction of the AdS bulk can emerge from quantum entanglement in the dual conformal field theory (CFT)~\cite{h,hh}. In this context, the holographic entanglement entropy associated with a boundary subregion $A$ is given by the area of an extremal surface $\Gamma_A$ in the bulk that is anchored on the boundary of $A$~\cite{r},
\begin{equation}
	S_A = \frac{\text{Area}(\Gamma_A)}{4G},
\end{equation}
where $G$ is Newton's constant. This quantity generalizes the Bekenstein-Hawking entropy to arbitrary boundary subregions and is known as the Ryu-Takayanagi entropy.

This result naturally motivates a fundamental question: can the time coordinate, in analogy with spatial directions, also emerge from quantum entanglement? In order to address this question, a new physical concept has been introduced, known as \emph{timelike entanglement entropy}~\cite{t}. This quantity is explicitly associated with temporal quantum correlations and provides a framework for probing the role of entanglement in the emergence of time.
As discussed in the introduction, holographic timelike entanglement entropy can be interpreted as a \emph{pseudoentropy}. It is defined by~\cite{Ca}
\begin{equation}
	S_A = -\mathrm{Tr}\!\left[\rho_A \log \rho_A \right],
\end{equation}
where $\rho_A$ denotes the reduced density matrix associated with a temporal partition. The reduced density matrix is given by
\begin{equation}
	\rho_A = \mathrm{Tr}_B \left[ \frac{|\psi\rangle \langle \varphi|}{\langle \varphi | \psi \rangle} \right],
\end{equation}
with $|\psi\rangle$ and $|\varphi\rangle$ representing two distinct quantum states defined on the total Hilbert space, which factorizes as $\mathcal{H} = \mathcal{H}_A \otimes \mathcal{H}_B$.

To study holographic timelike entanglement entropy, we consider a two-dimensional conformal field theory and define a subsystem $A$ by an interval $I = MN$, with endpoints $M = (t_M, x_M)$ and $N = (t_N, x_N)$. The von Neumann entropy of this subsystem can be expressed in terms of the geodesic distance between $M$ and $N$, the ultraviolet cutoff $\epsilon$, and the central charge $C$ of the theory~\cite{t},
\begin{equation}
	S_A = \frac{C}{3} \log \left( \frac{\sqrt{(x_M - x_N)^2 - (t_M - t_N)^2}}{\epsilon} \right).
\end{equation}

For the timelike geodesic, where $x_M = x_N$, the entropy becomes
\begin{equation}
	S_A = \frac{C}{3} \log \left( \frac{t_M - t_N}{\epsilon} \right) + \frac{\pi C}{6} i.
\end{equation}
This quantity defines the timelike entanglement entropy. Unlike the spacelike case, it is generally complex, with the imaginary contribution arising directly from the timelike nature of the geodesic. This behavior indicates that the timelike coordinate, similar to spatial coordinates, can be regarded as emerging from quantum entanglement in the boundary theory. Accordingly, these results support the view that spacetime itself is not fundamental but emerges from underlying quantum correlations. Further details and related discussions can be found in Refs.~\cite{t,11,13,14,15,16,17,18,19}.

\section{Hawking Radiation Entropy}\label{SecIII}

Hawking radiation \cite{Hawking1976} was originally introduced in the context of black hole evaporation, providing a mechanism by which black holes lose mass. In the semiclassical description, this process appears to violate the unitarity principle of quantum mechanics, since the entanglement entropy of the radiation does not follow the Page curve. Instead, the entropy increases monotonically with time, $S \propto t$, suggesting the presence of information loss. In this context, we introduce the concept of the \emph{timelike entanglement entropy} of Hawking radiation. 

\subsection{Schwarzschild black holes}

We first consider the Schwarzschild solution as a simple and tractable black hole model in order to illustrate this concept and examine its fundamental properties.
The spacetime geometry of a Schwarzschild black hole is described by the metric
\begin{equation}
	ds^2 = - f(r)\, dt^2 + \frac{dr^2}{f(r)} + r^2 d\theta^2 + r^2 \sin^2\theta\, d\phi^2 ,
\end{equation}
where the metric function $f(r)$ is given by
\begin{equation}
	f(r) = 1 - \frac{r_h}{r}.
\end{equation}
Here  $r_h = 2MG$ is the event horizon radius,  $M$ is the black hole mass and $G$ is Newton’s gravitational constant.

Hawking radiation is emitted in the asymptotic region far from the black hole and is characterized by a thermal spectrum with the Hawking temperature
\begin{equation}
	T_H = \frac{\kappa}{2\pi} = \frac{f'(r_h)}{4\pi} = \frac{1}{4\pi r_h},
\end{equation} 
where $\kappa$ is the surface gravity at the event horizon, and the prime denotes differentiation with respect to $r$.

In our analysis, we adopt the $s$-wave approximation and neglect the angular degrees of freedom at large distances, allowing the four-dimensional spacetime to be effectively reduced to a two-dimensional spacetime with coordinates $(t,r)$~\cite{112}. Within this effective description, we study the entanglement entropy of Hawking radiation.

We consider a region of Hawking radiation defined by the interval $I = AB$, where the endpoints are $A = (t_a, x_a)$ and $B = (t_b, x_b)$. In the framework of the AdS/CFT correspondence, the von Neumann entropy of this radiation region, denoted by $S_R$, is identified with the entropy of a conformal field theory defined on the same interval. This relation can be written as~\cite{Li,CT}
\begin{equation}
	S_R = S_{CFT} = \frac{C}{3} \log \left( \frac{d(A,B)}{\epsilon} \right),
\end{equation}
where $C$ is the central charge of the conformal field theory, $d(A,B)$ denotes the geodesic distance between the points $A$ and $B$, and $\epsilon$ is an ultraviolet cutoff.

For a timelike region of Hawking radiation, characterized by $x_a = x_b$, we evaluate the entanglement entropy using an analytic continuation in time. In particular, we perform a Wick rotation $t \to i \tau$, under which the Lorentzian time coordinate is continued to Euclidean time. Under this transformation, the Schwarzschild metric takes the form~\cite{B22}
\begin{equation}
	ds^2_E = f(r)\, d\tau^2 + \frac{dr^2}{f(r)} + r^2 d\theta^2 + r^2 \sin^2\theta\, d\phi^2 .
\end{equation}

In the Euclidean geometry, the Kruskal coordinates are defined as
\begin{equation}
	\label{A1}
	U = - e^{-\kappa ( i \tau - r_* )}, \qquad 
	V = e^{\kappa ( i \tau + r_* )},
\end{equation}
where $r_*$ denotes the tortoise coordinate, given by~\cite{Li}
\begin{equation}
	r_*(a) = \int_{0}^{a} \frac{dr}{f(r)} .
\end{equation}

The geodesic distance between the points $A$ and $B$ can be expressed in terms of the Kruskal coordinates as~\cite{Li}
\begin{equation}
	d^2(A,B) = \Omega^2(x_a)\,\bigl(U(B) - U(A)\bigr)\bigl(V(A) - V(B)\bigr),
\end{equation}
with the conformal factor
\begin{equation}
	\label{A11}
	\Omega^2(x_a) = \frac{1}{\kappa^2 e^{2\kappa r_*(x_a)}} .
\end{equation}

Using Eqs.~\eqref{A1}–\eqref{A11}, the geodesic distance for a timelike separation is found to be
\begin{equation}
	d^2(A,B) = \frac{4}{\kappa^2}\,\sin^2\!\left(\frac{\kappa t}{2}\right),
\end{equation}
where $t = \tau_a - \tau_b$. The corresponding timelike entanglement entropy of Hawking radiation is therefore given by
\begin{equation}
	S_R(t) = S_{CFT}(t) = \frac{C}{6}\,\log\!\left( \frac{4\,\sin^2(\kappa t / 2)}{\kappa^2 \epsilon^2} \right).
\end{equation}

The time at which the entanglement entropy reaches its maximum is identified with the Page time $t_P$. It is determined by the condition
\begin{equation}
	\label{pt}
	\left. \frac{\partial S_R(t)}{\partial t} \right|_{t = t_P} = 0 .
\end{equation}
Solving this equation yields $	t_P = \pi/\kappa$.
At the Page time, the entropy of the Hawking radiation equals the Bekenstein-Hawking entropy of the black hole, given by $S_{BH} = \pi r_h^2 / G$~\cite{i}. Imposing this condition allows us to determine the ultraviolet cutoff, which takes the form
\begin{equation}
	\epsilon = \frac{2}{\kappa}\, \exp\!\left(-\frac{3\pi r_h^2}{C G}\right).
\end{equation}

\begin{figure}
	\centering
	\includegraphics[width=0.75\linewidth]{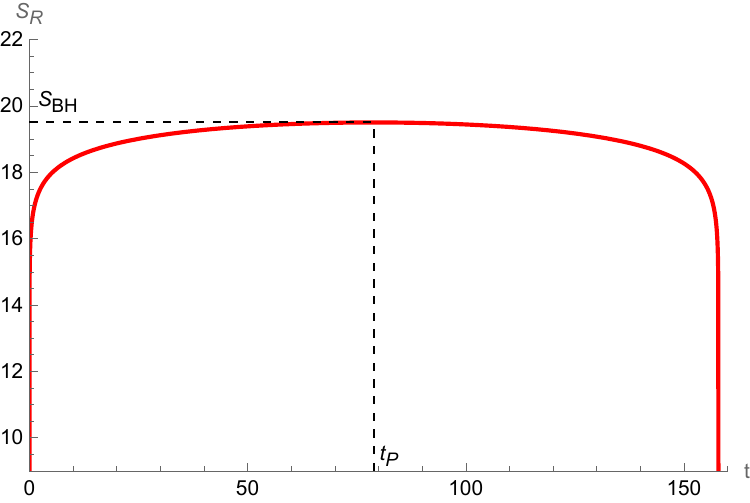}
	\caption{Timelike entanglement entropy of Hawking radiation as a function of time.}
	\label{f1}
\end{figure}

\begin{figure}
	\centering
	\includegraphics[width=0.75\linewidth]{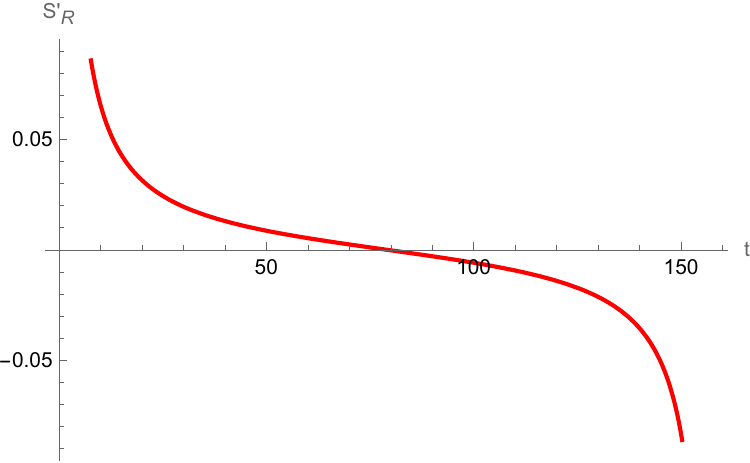}
	\caption{Derivative of the timelike entanglement entropy of Hawking radiation with respect to time, as a function of time.}
	\label{f2}
\end{figure}

Figures~\ref{f1} and~\ref{f2} show the time evolution of the timelike entanglement entropy and its time derivative, respectively. These results exhibit two distinct phases. Before the Page time, the entropy increases monotonically, while after the Page time it decreases. The maximum value of the entropy coincides with the Bekenstein-Hawking entropy of the black hole.
In the first phase, prior to the Page time, the Hawking radiation is predominantly entangled with the black hole. At the Page time, an information balance is reached, characterized by the equality between the entropy of the black hole and that of the Hawking radiation. In the second phase, after the Page time, information is gradually recovered in the Hawking radiation trough timelike entanglement.

Consequently, the timelike entanglement entropy reproduces the Page curve and remains consistent with unitary evolution. This behavior indicates that information is not lost during black hole evaporation, but is instead encoded in the timelike entanglement structure of Hawking radiation. The notion of timelike entanglement therefore provides a possible resolution of the black hole information paradox.

On the other hand, the expression for the Page time, $t_P \propto 1/\kappa$, shows that it is directly controlled by the surface gravity of the black hole. Since the surface gravity is inversely proportional to the event horizon radius, the Page time depends on the size of the black hole. As a result, the black hole radius determines the location of the maximum of the timelike entanglement entropy, which marks the onset of information retrieval in the Hawking radiation.
In particular, larger black holes, which have smaller surface gravity, exhibit longer Page times, implying that information recovery takes longer than for smaller black holes.
This dependence highlights a nontrivial interplay between quantum mechanics, gravity, and thermodynamics in the process of black hole evaporation.

We extend our analysis to the full time evolution of the system, without restricting attention to the first period of the timelike entanglement entropy. Owing to its functional form, the timelike entanglement entropy exhibits a periodic behavior in time. As a consequence, the Page time is not unique, but instead occurs at a discrete set of times determined by all solutions of Eq.~\eqref{pt},
\begin{equation}
	t_{P_n} = \frac{\pi}{\kappa} + \frac{2n\pi}{\kappa},
\end{equation}
where $n \in \mathbb{Z}_{\ge 0}$ labels the successive periods of the timelike entanglement entropy. At each of these times, the timelike entanglement entropy coincides with the black hole entropy,
\begin{equation}
	S_R(t_{P_n}) = S_{BH}(t_{P_n}).
\end{equation}

At late times, corresponding to $t_{P_n} \to \infty$, the black hole completely evaporates and ceases to exist. In this limit, the timelike entanglement entropy vanishes, and its periodic behavior asymptotically settles at zero. This regime represents the final stage of information recovery. Within each period of the evolution, information is effectively lost during the phase in which the timelike entanglement entropy increases, and it is recovered during the phase in which the entropy decreases. This alternating process continues until the black hole fully disappears. At that point, the timelike entanglement entropy remains zero, signaling the complete restoration of a pure quantum state.
Consequently, the full time evolution respects unitarity, and the black hole information paradox is resolved within the framework of timelike entanglement.

\subsection{Reissner-Nordstr\"om (RN) black holes}

For Reissner-Nordstr\"om (RN) black holes, characterized by mass $M$ and electric charge $Q$, the spacetime metric possesses two horizons, namely, an outer (event) horizon $r_+$ and an inner (Cauchy) horizon $r_-$, located at
\begin{equation}
	r_\pm = M \pm \sqrt{M^2 - Q^2}.
\end{equation}

The surface gravity at the outer horizon, which controls the Hawking temperature and the periodicity of the Euclidean time, is
\begin{equation}
	\kappa_{\rm RN} = \frac{r_+ - r_-}{2 r_+^2} 
	= \frac{\sqrt{M^2 - Q^2}}{(M + \sqrt{M^2 - Q^2})^2}.
\end{equation}
Since $\kappa_{\rm RN}$ decreases monotonically with increasing charge $Q$, the corresponding Euclidean time periodicity, $\beta_{\rm RN} = 2\pi/\kappa_{\rm RN}$, grows. In the context of timelike entanglement entropy, this implies that the characteristic oscillation cycles of the entropy become longer as the black hole approaches extremality. Physically, the slower oscillations reflect the decreased Hawking temperature and reduced energy flux from the black hole.

The timelike Page times for successive entanglement recurrences can be written as
\begin{equation}
	t_{P_n}^{\rm RN} = \frac{\pi}{\kappa_{\rm RN}} + \frac{2 \pi n}{\kappa_{\rm RN}}, \quad n \in \mathbb{Z}_{\ge 0}.
\end{equation}
Because $\kappa_{\rm RN}$ decreases with increasing charge, $t_{P_n}^{\rm RN}$ increases, implying that the information recovery from the black hole slows down. This is consistent with the thermodynamic stability of RN black holes: charged black holes radiate more slowly and are less prone to rapid evaporation compared to their Schwarzschild counterparts~\cite{Gibbons:1977mu}.

In the extremal limit, $Q \to M$, the horizons coincide, $r_+ \to r_-$, and the surface gravity vanishes, $\kappa_{\rm RN} \to 0$.
Consequently, the period of the timelike entanglement entropy diverges, and the pseudoentropy becomes effectively frozen. This reflects the zero-temperature nature of extremal RN black holes, whose near-horizon geometry develops an infinite AdS$_2$ throat, a feature central to studies of quantum gravity and holography~\cite{Strominger:1998yg,Sen:2005wa}.

Thus, these observations indicate that the timelike Page curve may serve as a sensitive probe of near-extremal dynamics, encoding both thermodynamic slowdowns and geometric features of the black hole interior. In particular, the divergence of the timelike Page period near extremality provides a direct link between entanglement evolution and the deep gravitational structure of charged black holes.

\subsection{Static higher-dimensional and braneworld black holes}

\subsubsection{Static higher-dimensional black holes}

In $D>4$ spacetime dimensions, the natural generalization of the Schwarzschild solution is the Schwarzschild--Tangherlini metric~\cite{Tangherlini:1963bw}, which describes a static, spherically symmetric black hole. The line element reads
\begin{equation}
	ds^2 = - f_D(r) \, dt^2 + f_D(r)^{-1} \, dr^2 + r^2 \, d\Omega_{D-2}^2,
\end{equation}
where \(d\Omega_{D-2}^2\) is the metric on the unit \((D-2)\)-sphere and the metric function is
\begin{equation}
	f_D(r) = 1 - \left( \frac{r_h}{r} \right)^{D-3}.
\end{equation}
Here, \(r_h\) is the horizon radius, related to the ADM mass \(M\) via
\begin{equation}
	M = \frac{(D-2)\,\Omega_{D-2}}{16\pi G} \, r_h^{D-3}.
\end{equation}

The surface gravity, \(\kappa_D\), is obtained from the general formula
\begin{equation}
	\kappa_D = \frac{1}{2} f_D'(r_h) = \frac{D-3}{2 r_h},
\end{equation}
where the prime denotes differentiation with respect to $r$.
This shows that the surface gravity, and hence the Hawking temperature, \(T_H = \kappa_D / 2\pi\), grows with dimensionality \(D\) for a fixed horizon radius. Consequently, the thermal timescale of entanglement recurrences decreases in higher dimensions, leading to shorter timelike Page cycles.

By analogy with the 4D Schwarzschild case, the timelike Page times associated with entanglement recurrences are
\begin{equation}
	t_{P_n}^{(D)} = \frac{\pi}{\kappa_D} + \frac{2\pi n}{\kappa_D}, \quad n \in \mathbb{Z}_{\ge 0}.
\end{equation}

The first term \(\pi / \kappa_D\) corresponds to the half-period of the pseudoentropy oscillation, while the second term gives successive recurrences. The higher-dimensional scaling of \(\kappa_D\) implies that \(t_{P_n}^{(D)}\) decreases with increasing \(D\) for fixed \(r_h\), signaling faster information recovery in higher-dimensional spacetimes.

\subsubsection{Braneworld and extra-dimensional corrections}

In braneworld scenarios such as the Randall-Sundrum model~\cite{Randall:1999vf}, the effective four-dimensional black hole inherits corrections from the bulk spacetime. These include Kaluza-Klein modes, higher-curvature terms, or tidal contributions from extra dimensions. In practice, these corrections modify the effective horizon radius \(r_h^{\rm eff}\) and surface gravity \(\kappa_D^{\rm eff}\)
\begin{equation}
	\kappa_D^{\rm eff} = \frac{D-3}{2 r_h^{\rm eff}} + \delta\kappa,
\end{equation}
where \(\delta \kappa\) encodes the influence of extra-dimensional physics.
Consequently, the timelike Page times are also modified
\begin{equation}
	t_{P_n}^{\rm eff} = \frac{\pi}{\kappa_D^{\rm eff}} + \frac{2\pi n}{\kappa_D^{\rm eff}}, \quad n \in \mathbb{Z}_{\ge 0}.
\end{equation}

Such deviations in \(t_{P_n}\) provide a holographically motivated diagnostic of extra-dimensional effects, as they directly probe changes in the effective surface gravity and entanglement recurrence times.

\subsection{Rotating Kerr black holes}

Kerr black holes, characterized by mass $M$ and angular momentum per unit mass $a = J/M$, introduce qualitatively new features into the temporal entanglement structure due to their rotation. The spacetime possesses an outer  horizon $r_+$ and an inner  horizon $r_-$ given by
\begin{equation}
	r_\pm = M \pm \sqrt{M^2 - a^2}.
\end{equation}

The surface gravity at the outer horizon is
\begin{equation}
	\kappa_{\rm Kerr} = \frac{r_+ - r_-}{2(r_+^2 + a^2)},
\end{equation}
and the horizon rotates with angular velocity
\begin{equation}
	\Omega_H = \frac{a}{r_+^2 + a^2}.
\end{equation}

Unlike spherically symmetric spacetimes, the Kerr geometry is only axisymmetric, introducing explicit angular dependence into timelike geodesic distances. Consequently, the timelike entanglement entropy depends on both $\kappa_{\rm Kerr}$ and $\Omega_H$, and the resulting Page curves are generally quasi-periodic rather than strictly periodic. Physically, this quasi-periodicity can be understood as a beat phenomenon arising from the interference between the thermal timescale $2\pi/\kappa_{\rm Kerr}$ and the rotational timescale $2\pi/\Omega_H$, analogous to oscillatory behaviors in rotating thermal CFTs~\cite{Castro:2013kea}.

The analytic continuation of Kerr spacetime to Euclidean signature introduces a complex metric with nontrivial identifications in both Euclidean time $\tau$ and the azimuthal angle $\phi$~\cite{Gibbons:1976ue}
\begin{equation}
	(\tau, \phi) \sim (\tau + \beta_{\rm Kerr}, \phi - i \beta_{\rm Kerr} \Omega_H),
\end{equation}
where $\beta_{\rm Kerr} = 2\pi/\kappa_{\rm Kerr}$. This structure implies that the imaginary part of the pseudoentropy may encode not only causal separation but also rotational phases, in a manner analogous to chemical potentials in grand canonical ensembles.  

The timelike Page times are therefore influenced by both the thermal and rotational scales. The thermal contribution is
\begin{equation}
	t_{P_n}^{\rm thermal} = \frac{\pi}{\kappa_{\rm Kerr}} + \frac{2 \pi n}{\kappa_{\rm Kerr}}, \quad n \in \mathbb{Z}_{\ge 0},
\end{equation}
while successive recurrences also acquire a rotational phase
\begin{equation}
	\phi_n = n \, 2 \pi \frac{\Omega_H}{\kappa_{\rm Kerr}}.
\end{equation}

Combining these effects, the quasi-periodic timelike Page times can be expressed as
\begin{equation}
	t_{P_n}^{\rm Kerr} \simeq \frac{\pi}{\kappa_{\rm Kerr}} + \frac{2 \pi n}{\kappa_{\rm Kerr}} + f\!\left(n, \frac{\Omega_H}{\kappa_{\rm Kerr}}\right),
\end{equation}
where $f(n, \Omega_H/\kappa_{\rm Kerr})$ captures the interference between the thermal and rotational scales. A simple approximation for the leading-order modulation is
\begin{equation}
	f\!\left(n, \frac{\Omega_H}{\kappa_{\rm Kerr}}\right) \sim \frac{1}{\kappa_{\rm Kerr}} \, \sin\!\Big( 2 \pi n \frac{\Omega_H}{\kappa_{\rm Kerr}} \Big),
\end{equation}
producing modulated recurrences rather than strictly periodic ones.

In the extremal limit $a \to M$, the surface gravity vanishes, $\kappa_{\rm Kerr} \to 0$, while the horizon angular velocity approaches its maximal value, $\Omega_H \to 1/(2M)$. Consequently, the thermal recurrence times diverge,
\begin{equation}
	t_{P_n}^{\rm thermal} \to \infty,
\end{equation}
but the rotational modulation remains finite. This leads to effectively frozen thermal evolution but persistent angular variation in the pseudoentropy, reflecting the structure of the infinite near-horizon $\mathrm{AdS}_2 \times S^1$ throat geometry.

Overall, the timelike Page times for Kerr black holes encode the competition between the thermal scale $2\pi/\kappa_{\rm Kerr}$ and the rotational scale $2\pi/\Omega_H$. Their interplay produces quasi-periodic entanglement recurrences, providing a sensitive probe of near-horizon dynamics and the microscopic structure of rotating black holes.

\subsection{Higher‑dimensional rotating black holes}

In higher than four spacetime dimensions \(D>4\), black hole solutions exhibit richer horizon structure and multiple independent rotation planes. The canonical generalization of the Kerr solution in \(D\) dimensions is the Myers-Perry (MP) family of rotating black holes~\cite{Myers:1986un}, which reduces to the Tangherlini solution~\cite{Tangherlini:1963bw} in the non‑rotating limit. In this section, we derive analytic expressions for the timelike Page times in higher dimensions, emphasizing how dimensionality and rotation interplay in the temporal entanglement structure.

The \(D\)‑dimensional Myers-Perry metric with a single nonzero rotation parameter \(a\) takes the form
\begin{align}
	ds^2 &= -dt^2 + \frac{\mu r^{5-D}}{\Sigma}\left(dt - a \sin^2\theta\,d\phi\right)^2 + \frac{\Sigma}{\Delta}\,dr^2 + \Sigma\,d\theta^2\nonumber\\
	&\quad + \left(r^2 + a^2\right)\sin^2\theta\,d\phi^2 + r^2\cos^2\theta\,d\Omega_{D-4}^2,
\end{align}
where
\begin{equation}
	\Delta = r^2 + a^2 - \mu r^{5-D},\qquad \Sigma = r^2 + a^2\cos^2\theta,
\end{equation}
and \(d\Omega_{D-4}^2\) is the round metric on \(S^{D-4}\). The constant \(\mu\) is related to the ADM mass \(M\) by
\begin{equation}
	M = \frac{(D-2)\,\Omega_{D-2}}{16\pi G}\,\mu,
\end{equation}
with \(\Omega_{D-2}\) the volume of the unit \((D-2)\)‑sphere.

The event horizon is defined by the largest real root \(r_+\) of \(\Delta(r_+)=0\). In contrast to four dimensions, rotating horizons in higher \(D\) can admit multiple rotation planes; here we focus on a single rotation parameter for clarity.


The Killing vector that generates the horizon is
\begin{equation}
	\xi = \partial_t + \Omega_H\,\partial_\phi,
\end{equation}
where the horizon angular velocity is
\begin{equation}
	\Omega_H = \frac{a}{r_+^2 + a^2}.
\end{equation}

The surface gravity \(\kappa_D\) associated with \(\xi\) is obtained from
\begin{equation}
	\kappa_D^2 = -\tfrac{1}{2}\,\big(\nabla^\mu \xi^\nu\big)\big(\nabla_\mu \xi_\nu\big)\big|_{r=r_+}.
\end{equation}
Direct evaluation yields
\begin{equation}
	\kappa_D = \frac{(D-3)r_+^2 + (D-5)a^2}{2 r_+\,(r_+^2 + a^2)},
\end{equation}
which reduces smoothly to the Tangherlini surface gravity in the non‑rotating limit \(a\to 0\),
\begin{equation}
	\kappa_{\rm Tangherlini} = \frac{D-3}{2\,r_h}.
\end{equation}


To implement thermal boundary conditions, we Wick rotate \(t\to -i\tau\). In higher dimensions, as in four dimensions, the Euclidean section of a rotating black hole generally becomes complex due to cross terms mixing \(\tau\) and \(\phi\). Regularity at the horizon requires the simultaneous identification
\begin{equation}
	(\tau,\phi)\sim(\tau+\beta_D,\phi-i\,\beta_D\,\Omega_H),
\end{equation}
where $\beta_D = 2\pi/\kappa_D$ is the inverse Hawking temperature. These identifications generalize the \(D=4\) Kerr periodicities by encoding thermal and rotational cycles together. Crucially, the presence of multiple angular directions in general \(D\) adds independent chemical potentials \(i\beta_D\,\Omega_{H_i}\) for each rotation plane, reflecting grand canonical ensembles with multiple angular potentials.


The timelike Page times inherit both the thermal timescale and the rotational modulation from the Euclidean identifications. The purely thermal contribution generalizes straightforwardly
\begin{equation}
	t_{P_n}^{\rm thermal} = \frac{\pi}{\kappa_D} + \frac{2\pi n}{\kappa_D},\qquad n\in\mathbb{Z}_{\ge0},
\end{equation}
where higher \(D\) typically increases \(\kappa_D\) for fixed horizon radius, leading to shorter Page cycles relative to the four‑dimensional case. This reflects the fact that the effective temperature of a higher‑dimensional black hole is generally higher for fixed geometric scale, accelerating information recovery.

Rotation enters through successive angular phases
\begin{equation}
	\phi_n = n\,2\pi\,\frac{\Omega_H}{\kappa_D},
\end{equation}
with additional phases for each independent rotation plane if present. The combined recurrence times thus take a quasi‑periodic form,
\begin{equation}
	t_{P_n}^{\rm MP} \simeq \frac{\pi}{\kappa_D} + \frac{2\pi n}{\kappa_D} + f\!\left(n,\frac{\Omega_H}{\kappa_D}\right),
\end{equation}
where the modulation function \(f\) captures interference between thermal and rotational scales. As before, a simple leading approximation that encodes the dominant beat behavior is
\begin{equation}
	f\!\left(n,\frac{\Omega_H}{\kappa_D}\right)\sim\frac{1}{\kappa_D}\,\sin\!\Big(2\pi\,n\,\frac{\Omega_H}{\kappa_D}\Big),
\end{equation}
producing oscillatory recurrences that deviate from strict periodicity.


In the extremal limit, where one or more rotation parameters saturate an upper bound (\(a \to a_{\rm ext}(M,D) \gg r_+\)), this similarly drives \(\kappa_D \to 0\) in $5D$. However, for $D \ge 6$, no extremal limit exists.

The extremal limit in $5D$ corresponds to ultra-spinning black holes, while the corresponding horizon angular velocity remains finite. Consequently
\begin{equation}
	t_{P_n}^{\rm thermal}\to\infty,
\end{equation}
reflecting effectively frozen thermal oscillations, whereas the rotational modulation persists. In the presence of multiple rotation planes, each angular potential contributes an independent modulation channel, leading to richer quasi‑periodic structures in the timelike Page curve.

\section{Discussion and Conclusion}\label{Sec:Conclusion}

In this work, we have introduced the concept of \emph{timelike entanglement entropy} as a tool to investigate the information content of Hawking radiation. By performing an analytic continuation of black hole spacetimes to Euclidean signature, we defined timelike correlations in the radiation and demonstrated that the associated entropy exhibits a periodic structure characterized by a sequence of \emph{timelike Page times}. At these times, the timelike entanglement entropy matches the Bekenstein--Hawking entropy of the black hole, signaling transitions between apparent information loss and recovery. Before each Page time, the entropy increases, reflecting the growth of entanglement due to Hawking emission, whereas after each Page time, the entropy decreases, indicating the retrieval of information. In the limit of complete evaporation, the timelike entropy and its periodicity vanish, leaving a pure final state and thus preserving unitarity throughout the evaporation process.

We have applied this framework across a broad range of black hole solutions. For Reissner-Nordstr\"om black holes, the surface gravity decreases with increasing charge, resulting in longer timelike Page cycles and slower information recovery, with extremal limits corresponding to effectively frozen timelike entropy and an infinite near-horizon $\mathrm{AdS}_2$ throat geometry. For static higher-dimensional black holes, the Schwarzschild-Tangherlini solution demonstrates that increasing spacetime dimensionality accelerates information recovery by increasing the surface gravity, while braneworld and Kaluza-Klein corrections alter the effective horizon radius and surface gravity, imprinting the presence of extra dimensions on the timelike Page curve. In the case of rotating black holes, both four-dimensional Kerr and higher-dimensional Myers-Perry solutions exhibit a rich interplay between thermal and rotational scales, producing quasi-periodic timelike Page times. The resulting beat-like modulation encodes information about angular momentum, extremality, and spacetime dimension, with thermal recurrences diverging in the extremal limit while rotational modulation persists. Multiple rotation planes in higher dimensions further enrich this quasi-periodic structure.

The timelike entanglement approach provides a conceptually distinct resolution to the black hole information paradox. Unlike the island formula, which relies on including regions inside the black hole via replica wormholes, or the firewall proposal, which violates the smoothness of the horizon, timelike entanglement encodes information entirely within the Hawking radiation itself. This mechanism preserves unitarity, maintains the smoothness of the horizon, and reproduces the qualitative features of the Page curve without invoking any assumptions about interior degrees of freedom or exotic structures.

These results point to several promising directions for future research. Extending our analysis to black holes with multiple rotation planes and more general higher-dimensional geometries could uncover richer quasi-periodic structures in timelike Page curves. Incorporating fully time-dependent or evaporating spacetimes would allow a more complete description of the dynamical flow of information, while connections with holographic duality and near-horizon $\mathrm{AdS}_2$ physics may provide deeper insight into the microscopic structure of black holes. Additionally, the sensitivity of timelike entanglement to rotation, dimensionality, and extra-dimensional effects suggests that it can serve as a diagnostic tool in both theoretical models and analogue gravity systems.

Overall, our study establishes timelike entanglement as a robust and conceptually transparent mechanism for information recovery in black hole evaporation. By unifying analyses across charged, rotating, and higher-dimensional black holes, this framework complements existing approaches such as the island formula and firewall scenarios, while offering a new perspective on the preservation of unitarity. These results open new avenues for exploring the quantum nature of gravity, the emergence of spacetime, and the fundamental role of temporal correlations in black hole thermodynamics.

   \acknowledgments{
   	YL gratefully acknowledges the support from the ``PhD-Associate Scholarship -- PASS'' grant provided by the National Center for Scientific and Technical Research in Morocco, under grant number 42 UMP2023.
   	FSNL acknowledges support from the Funda\c{c}\~{a}o para a Ci\^{e}ncia e a Tecnologia (FCT) Scientific Employment Stimulus contract with reference CEECINST/00032/2018, and funding through the research grants UID/04434/2025 and PTDC/FIS-AST/0054/2021.}


\begin{thebibliography}{9}

\bibitem{RovelliLQG}
C.~Rovelli,
``Loop Quantum Gravity,''  Living Rev. Rel. \textbf{11} (2008), 5
[arXiv:gr-qc/9710008].


\bibitem{PolchinskiBook}
M.~Grana,
``Flux Compactifications in String Theory: A Comprehensive Review,''  Phys. Rept. \textbf{423} (2006), 91
[arXiv:hep-th/0509003].

\bibitem{Maldacena1997}
J.~M.~Maldacena,
``The Large N Limit of Superconformal Field Theories and Supergravity,''  Adv. Theor. Math. Phys. \textbf{2} (1998), 231
[arXiv:hep-th/9711200].

\bibitem{Hawking1975}
S.~W.~Hawking,
``Particle Creation by Black Holes,''  Commun. Math. Phys. \textbf{43} (1975), 199.




\bibitem{Hawking1976}
S.~W.~Hawking,
``Breakdown of Predictability in Gravitational Collapse,''  Phys. Rev. D \textbf{14} (1976), 2460.

\bibitem{Page1993a}
D.~N.~Page,
``Information in Black Hole Radiation,''  Phys. Rev. Lett. \textbf{71} (1993), 3743
[arXiv:hep-th/9306083].

\bibitem{Page1993b}
D.~N.~Page,
``Time Dependence of Hawking Radiation Entropy,''  JCAP \textbf{09} (2013), 028
[arXiv:1301.4995 [hep-th]].

\bibitem{AMPS2013}
A.~Almheiri, D.~Marolf, J.~Polchinski, J.~Sully,
``Black Holes: Complementarity or Firewalls?,''  JHEP \textbf{02} (2013), 062
[arXiv:1207.3123 [hep-th]].




\bibitem{Penington2019}
G.~Penington,
``Entanglement Wedge Reconstruction and the Information Paradox,''  JHEP \textbf{09} (2020), 002
[arXiv:1905.08255 [hep-th]].

\bibitem{Almheiri2019}
A.~Almheiri, N.~Engelhardt, D.~Marolf, H.~Maxfield,
``The entropy of bulk quantum fields and the entanglement wedge of an evaporating black hole,''  JHEP \textbf{12} (2019), 063
[arXiv:1905.08762 [hep-th]].

\bibitem{Coffman2000}
V.~Coffman, J.~Kundu, W.~K.~Wootters,
``Distributed Entanglement,''  Phys. Rev. A \textbf{61} (2000), 052306
[arXiv:quant-ph/9907047].

\bibitem{i}
A.~Almheiri, T.~Hartman, J.~Maldacena, E.~Shaghoulian, A.~Tajdini,
``The entropy of Hawking radiation,''  Rev. Mod. Phys. \textbf{93} (2021), 035002
[arXiv:2006.06872 [hep-th]].

\bibitem{iii}
K.~Hashimoto, N.~Iizuka, Y.~Matsuo,
``Islands in Schwarzschild black holes,''  JHEP \textbf{06} (2020), 085
[arXiv:2004.05863 [hep-th]].



\bibitem{ss}
P.~X.~Hao, T.~Kawamoto, S.~M.~Ruan, T.~Takayanagi,
``Non-extremal island in de Sitter gravity,''  JHEP \textbf{03} (2025), 004
[arXiv:2407.21617 [hep-th]].

\bibitem{sss}
K.~Goswami, K.~Narayan,  
``Small Schwarzschild de Sitter black holes, quantum extremal surfaces and islands,''  JHEP \textbf{10} (2022), 031
[arXiv:2207.10724 [hep-th]].

\bibitem{ssss}
K.~Goswami, K.~Narayan,  
``Small Schwarzschild de Sitter black holes, the future boundary and islands,''  JHEP \textbf{05} (2024), 016
[arXiv:2312.05904 [hep-th]].

\bibitem{Lii}
R.~Li, X.~Wang, J.~Wang,
``Island may not save the information paradox of Liouville black holes,''  Phys. Rev. D \textbf{104} (2021), 106015
[arXiv:2105.03271 [hep-th]].

\bibitem{Oo}
S.~J.~Olson, T.~C.~Ralph,
``Extraction of timelike entanglement from the quantum vacuum,''  Phys. Rev. A \textbf{85} (2012), 012306
[arXiv:1101.2565 [quant-ph]].



\bibitem{AA}
M.~Nowakowski, E.~Cohen, P.~Horodecki,
``Entangled histories versus the two-state-vector formalism: Towards a better understanding of quantum temporal correlations,''  Phys. Rev. A \textbf{98} (2018), 032312
[arXiv:1803.11267 [quant-ph]].

\bibitem{DD}
A.~Jamiołkowski,
``Linear transformations which preserve trace and positive semidefiniteness of operators,''  Rep. Math. Phys. \textbf{3} (1972), 275.

\bibitem{EE}
M.-D.~Choi,
``Completely positive linear maps on complex matrices,''  Linear Algebra Appl. \textbf{10} (1975), 285.

\bibitem{J3}
M.~Jiang, S.~Luo, S.~Fu,
``Channel-state duality,''  Phys. Rev. A \textbf{87} (2013), 022310.

\bibitem{FF}
R.~P.~Feynman, F.~L.~Vernon Jr.,
``The theory of a general quantum system interacting with a linear dissipative system,''  Ann. Phys. \textbf{24} (1963), 118-173.

\bibitem{H3}
J.~Harper, A.~Mollabashi, T.~Takayanagi, Y.~Taki,
``Timelike entanglement entropy,''  JHEP \textbf{05} (2023), 123
[arXiv:2302.11695 [hep-th]].




\bibitem{r}
S.~Ryu and T.~Takayanagi,
``Holographic Derivation of Entanglement Entropy from the anti–de Sitter Space/Conformal Field Theory Correspondence,''  Phys. Rev. Lett. \textbf{96} (2006), 181602
[arXiv:hep-th/0603001].

\bibitem{t}
K.~Doi, J.~Harper, A.~Mollabashi, T.~Takayanagi, Y.~Taki,
``Pseudoentropy in dS/CFT and Timelike Entanglement Entropy,''  Phys. Rev. Lett. \textbf{130} (2023), 031601
[arXiv:2210.09457 [hep-th]].

\bibitem{h}
B.~Swingle,
``Entanglement Renormalization and Holography,''  Phys. Rev. D \textbf{86} (2012), 065007
[arXiv:0905.1317 [cond-mat.str-el]].

\bibitem{hh}
M.~Van Raamsdonk,
``Building Up Spacetime with Quantum Entanglement,''  Gen. Rel. Grav. \textbf{42} (2010), 2323--2329
[arXiv:1005.3035 [hep-th]].

\bibitem{Ca}
P.~Caputa, B.~Chen, T.~Takayanagi, T.~Tsuda,
``Thermal pseudo-entropy,''  JHEP \textbf{01} (2025), 003
[arXiv:2411.08948 [hep-th]].




\bibitem{11}
M.~P.~Heller, F.~Ori, A.~Serantes,
``Geometric Interpretation of Timelike Entanglement Entropy,''  Phys. Rev. Lett. \textbf{134} (2025), 131601
[arXiv:2408.15752 [hep-th]].



\bibitem{13}
Z.~Li, Z.~Q.~Xiao, R.~Q.~Yang,
``On Holographic Time-Like Entanglement Entropy,''  JHEP \textbf{04} (2023), 004
[arXiv:2211.14883 [hep-th]].

\bibitem{14}
X.~Jiang, P.~Wang, H.~Wu, H.~Yang,
``Timelike Entanglement Entropy and TT Deformation,''  Phys. Rev. D \textbf{108} (2023), 046004
[arXiv:2302.13872 [hep-th]].

\bibitem{15}
X.~Jiang, P.~Wang, H.~Wu, H.~Yang,
``Timelike Entanglement Entropy in dS$_3$/CFT$_2$,''  JHEP \textbf{08} (2023), 216
[arXiv:2304.10376 [hep-th]].

\bibitem{16}
M.~Afrasiar, J.~K.~Basak, D.~Giataganas,
``Timelike Entanglement Entropy and Phase Transitions in Non-Conformal Theories,''  JHEP \textbf{07} (2024), 243
[arXiv:2404.01393 [hep-th]].

\bibitem{17}
M.~Afrasiar, J.~K.~Basak, D.~Giataganas,
``Holographic Timelike Entanglement Entropy in Non-Relativistic Theories,''  JHEP \textbf{05} (2025), 205
[arXiv:2411.18514 [hep-th]].

\bibitem{18}
M.~P.~Heller, F.~Ori, A.~Serantes,
``Temporal Entanglement from Holographic Entanglement Entropy,''  Phys. Rev. X \textbf{15} (2025), 041022
[arXiv:2507.17847 [hep-th]].

\bibitem{19}
S.~S.~Jena, S.~Mahapatra,
``A Note on the Holographic Time-Like Entanglement Entropy in Lifshitz Theory,''  JHEP \textbf{01} (2025), 055
[arXiv:2410.00384 [hep-th]].



\bibitem{112}
C.~Z.~Guo, W.~C.~Gan, F.~W.~Shu,
``Page curves and entanglement islands for the step-function Vaidya model of evaporating black holes,''  JHEP \textbf{05} (2023), 042
[arXiv:2302.02379 [hep-th]].

\bibitem{CT}
P.~Calabrese, J.~Cardy,
``Entanglement entropy and quantum field theory,''  J. Stat. Mech. \textbf{0406} (2004), P06002
[arXiv:hep-th/0405152].

\bibitem{Li}
S.~Y.~Lin, M.~H.~Yu, X.~H.~Ge, L.~J.~Tian,
``Entanglement entropy, phase transition, and island rule for Reissner-Nordström-AdS black holes,''  Phys. Rev. D \textbf{110} (2024), 046008
[arXiv:2405.06873 [hep-th]].





\bibitem{B22}
E.~Battista, G.~Esposito,
``Geodesic motion in Euclidean Schwarzschild geometry,''  Eur. Phys. J. C \textbf{82} (2022), 1088
[arXiv:2202.03763 [gr-qc]].

\bibitem{Gibbons:1977mu}
G.~W.~Gibbons and S.~W.~Hawking,
``Cosmological Event Horizons, Thermodynamics, and Particle Creation,''  Phys. Rev. D \textbf{15} (1977), 2738-2751.

\bibitem{Strominger:1998yg}
A.~Strominger,
``AdS(2) quantum gravity and string theory,''  JHEP \textbf{01} (1999), 007
[arXiv:hep-th/9809027 [hep-th]].

\bibitem{Sen:2005wa}
A.~Sen,
``Black hole entropy function and the attractor mechanism in higher derivative gravity,''  JHEP \textbf{09} (2005), 038
[arXiv:hep-th/0506177 [hep-th]].




\bibitem{Tangherlini:1963bw}
F.~R.~Tangherlini,
``Schwarzschild field in n dimensions and the dimensionality of space problem,''  Nuovo Cim. \textbf{27} (1963), 636-651.



\bibitem{Castro:2013kea}
A.~Castro, J.~M.~Lapan, A.~Maloney and M.~J.~Rodriguez,
``Black Hole Monodromy and Conformal Field Theory,''  Phys. Rev. D \textbf{88} (2013), 044003
[arXiv:1303.0759 [hep-th]].

\bibitem{Gibbons:1976ue}
G.~W.~Gibbons and S.~W.~Hawking,
``Action Integrals and Partition Functions in Quantum Gravity,''  Phys. Rev. D \textbf{15} (1977), 2752-2756.



\bibitem{Randall:1999vf}
L.~Randall and R.~Sundrum,
``An Alternative to compactification,''  Phys. Rev. Lett. \textbf{83} (1999), 4690-4693
[arXiv:hep-th/9906064 [hep-th]].

\bibitem{turn0search7}
R.~Emparan and H.~S.~Reall,
``Black Holes in Higher Dimensions,''  Living Rev. Relativ. \textbf{11} (2008), 6
[arXiv:0801.3471 [hep-th]].

\bibitem{Myers:1986un}
R.~C.~Myers and M.~J.~Perry,
``Black Holes in Higher Dimensional Space-Times,''  Ann. Phys. \textbf{172} (1986), 304.



\end{thebibliography}
\end{document}